\documentclass[reprint,amsmath,amssymb,aps,prl,citeautoscript,footinbib]{revtex4-1}
\pdfoutput=1

\usepackage{graphicx}
\usepackage{epstopdf,epsfig}
\usepackage[scriptsize]{subfigure}
\usepackage[export]{adjustbox}
\usepackage{dcolumn}
\usepackage{bm}
\usepackage{marginnote}
\usepackage{changebar}
\usepackage{caption}
\usepackage[colorlinks]{hyperref}

\def\imagetop#1{\vtop{\null\scriptsize\hbox{#1}}}

\hypersetup{
    colorlinks=true,       
    linkcolor=blue,        
    citecolor=blue,        
    filecolor=black,       
    urlcolor=black,        
    bookmarks=false,
    pdffitwindow=true,
    pdfpagelayout=SinglePage,
    plainpages=false
}

\graphicspath{{figures-notes/}}

\renewcommand{\i}{\mathrm{i}}

\renewcommand{\d}{\mathrm{d}}

\newcommand{\be}{\begin{equation}}
\newcommand{\ee}{\end{equation}}
\newcommand{\ba}{\begin{eqnarray}}
\newcommand{\ea}{\end{eqnarray}}
\newcommand{\beg}{\begin{gather*}}
\newcommand{\eng}{\end{gather*}}

\newcommand{\bs}[1]{{\boldsymbol{#1}}}

\newcommand{\ins}[1]{{\mbox{\tiny #1}}}

\newcommand{\inds}[1]{{\scriptscriptstyle #1}}
\newcommand{\const}{\mbox{const}}

\begin{document}

\title{Lifshitz theory for a wedge}

\author{R. Krechetnikov}
\author{A. Zelnikov}
\email{zelnikov@ualberta.ca}
\affiliation{University of Alberta, Edmonton, Alberta, Canada T6G 2E1}

\date{\today}

\begin{abstract}
We develop the Lifshitz theory of van der Waals forces in a wedge of a dielectric material. The non-planar geometry of the problem requires determining point-wise distribution of stresses. The findings are relevant to a wide range of phenomena from crack propagation to contact line motion. First, the stresses prove to be anisotropic as opposed to the classical fluid mechanics treatment of the contact line problem. Second, the wedge configuration is always unstable with its angle tending either to collapse or unfold. The presented theory unequivocally demonstrates quantum nature of the forces dictating the wedge behavior, which cannot be accounted for with the classical methods.
\end{abstract}

\maketitle

Pressure in microscopically thin films is generally different from that in the macroscopic bulk due to the action of van der Waals forces \cite{Note1,van-der-Waals:1873}\footnotetext[1]{including orientation \cite{Keesom:1915}, induction \cite{Debye:1920}, and both non-retarded \cite{London:1930} and retarded \cite{Casimir:1948a,*Casimir:1948b} dispersion intermolecular} leading to disjoining pressure $\Pi_{\mathrm{D}}$. Originally it was calculated for pure substances \cite{Derjaguin:1934,*Derjaguin:1936,*Hamaker:1937}, under the assumption of additivity, via pair-wise summation of the attractive non-retarded part of the intermolecular potential $\varphi_{\mathrm{vdW}} \sim r^{-6}$:
\begin{align}\label{pressure:disjoining}
\Pi_{\mathrm{D}}(\ell) = - A_{\mathrm{H}} / 6 \pi \ell^{3},
\end{align}
where $\ell$ is the film thickness and $A_{\mathrm{H}}$ the Hamaker constant \cite{Note2}\footnotetext[2]{reflecting the strength of the molecular interaction between specific macroscopic bodies} specific to a given combination of substances in contact. Motivated by the discrepancy \cite{Note3} \footnotetext[3]{functional dependence $\ell^{-3}$ is still the same, though} between experiments \cite{Deryaguin:1954,*Deryaguin:1956,*Deryaguin:1960,*Tabor:1968} and the ``additive'' calculations, Lifshitz \cite{Lifshitz:1956} rigorously derived $\Pi_{\mathrm{D}}$ for dispersion forces \cite{Note4}\footnotetext[4]{called so because dispersion forces are due to the molecules polarizability, which in turn is related to the refractive index and thus dispersion} with QFT methods, thus recognizing their genuine quantum nature and non-additivity \cite{Note5,Farina:1999}\footnotetext[5]{unless rarefied, polarizability of a condensed matter may be vastly different from that of an individual molecule}, and naturally expressed $A_{\mathrm{H}}$ in terms of imaginary parts of the substances dielectric constants, in accordance with the fluctuation-dissipation theorem \cite{Note6,Rytov:1953}\footnotetext[6]{This fact is based on the Kramers-Kroning relation \cite{Lifshitz:1980}: their imaginary parts are always positive and determine the energy dissipation of the EM wave propagating in the medium.}.

\begin{figure}[h!]
\begin{tabular}{l l l l}
\hspace{-0.05in}\imagetop{(a)} & \hspace{-0.065in}\imagetop{\includegraphics[height=1.525in]{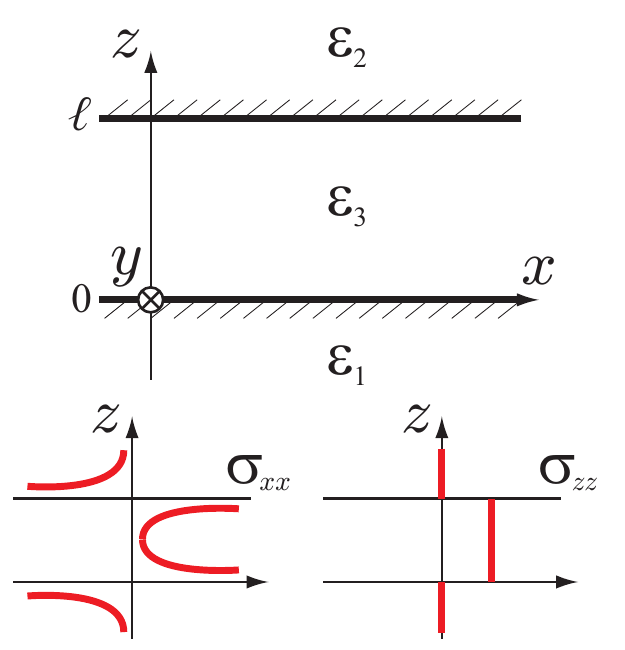}\label{fig:slit}} &
\hspace{0.05in}\imagetop{(b)} & \hspace{-0.065in}\imagetop{\includegraphics[height=1.525in]{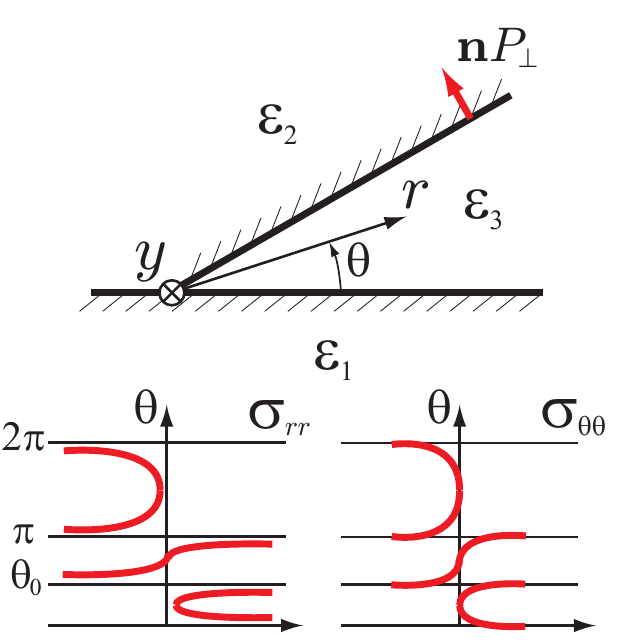}\label{fig:wedge}} \end{tabular}
\caption{Two basic configurations: (a) slit, (b) wedge.} \label{fig:configurations}
\end{figure}
This dispersive part of van der Waals forces is always present due to quantum fluctuations in the dielectric molecules' polarizability and plays a key role in a host of everyday phenomena such as adhesion, surface tension, adsorption, wetting, crack propagation in solids, to name a few \cite{Israelachvili:2011}. Since $\Pi_{\mathrm{D}}$ is prevalent at $\ell \lesssim 1 \, \mathrm{\mu m}$, it controls stability and wettability of liquid films. Due to constant value of this stress $\Pi_{\mathrm{D}}(\ell) \equiv -\sigma_{zz}$ across the film, it simply provides a jump in the total hydrodynamic pressure, cf. Fig.~\ref{fig:configurations}a. However, per \eqref{pressure:disjoining} $\Pi_{\mathrm{D}}$ diverges as $\ell \rightarrow 0$ and hence becomes invalid in modeling liquid films terminating at the substrate, which, in repetition of the history behind \eqref{pressure:disjoining}, has led to a number of attempts \cite{Note7}\footnotetext[7]{First, simple calculation of intermolecular potential for a wedge \cite{Miller:1974} brought some ideas about the forces driving wetting. Assuming that the intermolecular potential $\Phi$ is constant along the interface of the wedge with the contact angle $\theta_{0}$ at equilibrium, this led \cite{Hocking:1993} to the disjoining pressure in the small-slope $\ell_{x} \ll 1$ limit $\Pi_{\mathrm{D}}(\ell,\ell_{x}) \sim - \left(\theta_{0}^{4}-\ell_{x}^{4}\right)/\ell^{3}$, potentially regular at $\ell \rightarrow 0$. Requiring further that $\Phi$ is constant (minimized) not only at the interface, but also inside the wedge \cite{Wu:2004}, produced $\Pi_{\mathrm{D}}(\ell,\ell_{x},\ell_{xx}) \sim - \left(\theta_{0}^{4}-\ell_{x}^{4} + 2 \ell \ell_{x}^{2} \ell_{xx}\right)/\ell^{3}$, which, however, does not recover the planar film formula -- a fix for this issue was recently \cite{Dai:2008} suggested.} along the lines of the ``additive'' macroscopic theory \cite{Derjaguin:1934,*Derjaguin:1936,*Hamaker:1937} to generalize and regularize $\Pi_{\mathrm{D}}(\ell)$ for the wedge configuration, cf. Fig.~\ref{fig:configurations}b. These approaches not only ignored the non-additive nature of dispersion forces, which becomes especially important due to non-planar geometry of the problem thus affecting the stress distribution, but also relied upon the unjustified assumption of a uniform pressure across the wedge, in which anisotropy in the stress tensor near the interface is expected \cite{Berry:1971,Rusanov:2005,*Rusanov:2007}. Moreover, a priori it is clear that the singularity $\ell \rightarrow 0$ cannot be removed in the framework of van der Waals forces calculations due to intrinsic UV divergencies. Therefore, to properly account for van der Waals stresses in a wedge, one must generalize the Lifshitz film theory in order to determine local behavior of the energy-momentum stress tensor, in particular, to understand stability of the wedge configuration considered here in thermal equilibrium.

It must be noted that using Schwinger's source theory \cite{Schwinger:1978} rigorous calculations were recently done \cite{Brevik:1996,*Brevik:1998} of the Casimir effect for the wedge geometry, i.e. when the vacuum wedge region is bound by perfectly conducting walls and the dispersion forces are due to retarded potentials $\varphi_{\mathrm{vdW}} \sim r^{-7}$. Here, however, we are interested in dispersion forces in dielectrics and on shorter distances $\ell$ where retarded effects are no longer important, i.e. smaller than the wavelength $\lambda_{e}$ of the electromagnetic absorption peak, but larger than intermolecular separation $a_{0}$. In this non-retarded limit the Maxwell equations reduce to electrostatics with $A_{0}$ being the only non-zero component of the vector potential in the Coulomb gauge. Then, the Feynman propagator $G^F(t,\bs{r};t',\bs{r}')\equiv G_{00}^F(t,\bs{r};t',\bs{r}')$ is the vacuum expectation value (ground states corresponding to zero temperature) of the product of field operators $\i\langle \mathrm{T} \,\hat{A}_0(t,\bs{r})\hat{A}_0(t',\bs{r}') \rangle$; here the brackets $\langle\ldots\rangle$ denote averaging w.r.t. the ground state of the system and symbol $\mathrm{T}$ the chronological product, i.e. the operators following it are to be arranged from right to left in the order of increasing time. $G^F(t,\bs{r};t',\bs{r}')$ obeys
\be\label{GF}
\nabla_i(\varepsilon(\bs{r})\, \nabla^i) \, G^F(t,\bs{r};t',\bs{r}')=\delta(t,t')\delta(\bs{r},\bs{r}'),
\ee
in Planck units adopted throughout the Letter; here $\delta(\bs{r},\bs{r}')= {\delta(\bs{r}-\bs{r}') / \sqrt{-g}}$ and $\sqrt{-g} = r$ in the chosen cylindrical system of coordinates $\bs{r}=(y,r,\theta)$, cf. Fig.~\ref{fig:configurations}b.

When fluctuations are predominantly quantum \cite{Barash:1975} at the absorption frequency $\omega_{a}$, e.g. for water $T \le \omega_{a} = O(10^{3}) \, \mathrm{K}$, the Green's function of a macroscopic system (as ours) at non-zero temperatures differs from that at zero temperature only in that the averaging w.r.t. the ground state of a closed system is replaced by an averaging over the Gibbs distribution -- ensemble averages with thermal states at temperature $T \equiv \beta^{-1}$. The thermal Green's function $G^{\beta}(t,\bs{r};t',\bs{r}')$ can be obtained from the Feynman one using the Wick rotation -- substitution of $t = - \i t_\ins{E}$ in the Lorenzian Green's function to produce an Euclidean one $G^{\beta}(t_\ins{E},\bs{r},\bs{r}')= - \i \, G^F(\i t_\ins{E},\bs{r},\bs{r}')$, which is periodic in the Euclidean time $t_\ins{E}$ with period $\beta$; we also took into account the homogeneity of the Green's function in time and inhomogeneity in space in view of the presence of boundaries. Due to periodicity in Euclidean time, one can decompose $G^{\beta}(t_\ins{E},\bs{r},\bs{r}')$ and other related Green's functions in the Fourier time series:
\be
\label{FT:Green-thermal}
G^{\beta}(t_\ins{E},\bs{r},\bs{r}')
={1\over\beta}\sum_{n=-\infty}^{\infty}\widehat{G}^{\beta}(\zeta_n;\bs{r},\bs{r}') \, e^{\i\zeta_n t_\ins{E}};
\ee
here $\zeta_n={2\pi n / \beta}$ are the Matsubara frequencies and $\widehat{G}^{\beta}(\zeta_n;\bs{r},\bs{r}')$ the solution of the Fourier transformed equation \eqref{GF}, which in cylindrical coordinates reads
\be\label{eqG}
\nabla_i \nabla^i \widehat{G}^{\beta}(\zeta_n;\bs{r},\bs{r}')={1\over \varepsilon}{\delta(r-r')\delta(\theta-\theta')\delta(y-y')\over r},
\ee
where $\nabla_i \nabla^i = {1\over r}{\partial\over \partial{r}}r{\partial\over \partial{r}}+{1\over r^2}{\partial^2\over\partial\theta^2}+{\partial^2\over\partial y^2}$ is the Laplacian. At the boundary $\Sigma$ between two media $a$ and $b$ with the corresponding dielectric constants $\varepsilon_{a}$ and $\varepsilon_{b}$ one has to satisfy the standard boundary conditions
\begin{subequations}
\label{BCs:Green-thermal}
\begin{align}
\widehat{G}^{\beta}_{a}|_{\bs{r} \in \Sigma}&=\widehat{G}^{\beta}_{b}|_{\bs{r} \in \Sigma}, \\
\varepsilon_{a} n^i\nabla_i \,\widehat{G}^{\beta}_{a}|_{\bs{r} \in \Sigma}&=\varepsilon_{b} n^i\nabla_i \, \widehat{G}^{\beta}_{b}|_{\bs{r} \in \Sigma},
\end{align}
\end{subequations}
where $n^i$ is the normal vector to $\Sigma$. The constructed Green's function $\widehat{G}^{\beta}(\zeta_n;\bs{r},\bs{r}')$ also must be periodic in $\theta$ with period $2 \pi$ as per the problem statement, cf. Fig.~\ref{fig:configurations}b.

Applying the Fourier transform in the $y$-direction and, since electromagnetic surface waves decay exponentially away from the interface \cite{Lifshitz:1956}, the Kontorovich-Lebedev transform \cite{Kontorovich:1938,*Lebedev:1946}, commonly arising in diffraction problems on wedge-shaped domains, in the radial $r$-direction:
\begin{align}
\label{transform:Fourier-Kontorovich-Lebedev}
\widetilde{f}(\nu,k) = \int_{-\infty}^{\infty}{e^{-\i k y} \, \mathrm{d}k \int_{0}^{\infty}{\mathrm{d}r} f(r,y) \, K_{\i\nu}(k r) \, r^{-1}},
\end{align}
where $\nu$ has the meaning of a momentum in the $r$-direction and $K_{\i\nu}(k r)$ is the modified Bessel functions of the second kind, after rescaling $\widetilde{G}^{\beta} \rightarrow \Phi_{\nu} e^{-\i k y'} K_{\i\nu}(k r')$ we arrive at the boundary-value problem for the angular function $\Phi_{\nu}(\theta,\theta')$
\begin{align}
\label{eqPhi}
&\Big({\partial^2\over \partial{\theta}^2}-\nu^2\Big)\Phi_{\nu}(\theta,\theta') = {1\over{\varepsilon}} \, \delta(\theta-\theta'), \\
\begin{split}
&\Phi_{\nu}|_{\theta=\theta_0+0}=\Phi_{\nu}|_{\theta=\theta_0-0}, \, {\varepsilon_2}\partial_\theta\Phi_{\nu}|_{\theta=\theta_0+0}
={\varepsilon_3}\partial_\theta\Phi_{\nu}|_{\theta=\theta_0-0},
\\
&\Phi_{\nu}|_{\theta=\pi+0}=\Phi_{\nu}|_{\theta=\pi-0}, \, {\varepsilon_1}\partial_\theta\Phi_{\nu}|_{\theta=\pi+0}
={\varepsilon_2}\partial_\theta\Phi_{\nu}|_{\theta=\pi-0},
\\
&\Phi_{\nu}|_{\theta=+0}=\Phi_{\nu}|_{\theta=2\pi-0}, \, {\varepsilon_3}\partial_\theta\Phi_{\nu}|_{\theta=+0}
={\varepsilon_1}\partial_\theta\Phi_{\nu}|_{\theta=2\pi-0}.
\end{split} \nonumber
\end{align}
the solution of which takes the form
\be\label{Phi1}
\Phi_{\nu}(\theta,\theta')=
\begin{cases}
v e^{-\nu \theta}+d e^{-2\pi\nu} e^{\nu \theta}, &\pi<\theta<2\pi,\\
a e^{-\nu \theta}+ue^{\nu \theta}, &\theta_0< \theta< \pi,\\
b e^{\nu \theta}+ ce^{-\nu \theta}+{1\over 2\nu{\varepsilon_3}}e^{-\nu|\theta-\theta'|},&0< \theta<\theta_0.
\end{cases} \nonumber
\ee
The inverse transform corresponding to \eqref{transform:Fourier-Kontorovich-Lebedev} is given by
\begin{multline}
\label{transform-inverse:Fourier-Kontorovich-Lebedev}
\widehat{G}^{\beta}(\zeta_n;\bs{r},\bs{r}') = \frac{1}{\pi^{3}} \int_{-\infty}^{\infty}{e^{\i k (y-y')} \, \mathrm{d}k} \\ \int_{0}^{\infty}{\nu \, \sinh{(\pi \nu)} \, \mathrm{d}\nu \, \Phi_{\nu}(\theta,\theta') \, K_{\i\nu}(k r) \, K_{\i\nu}(k r')}.
\end{multline}

In the homogeneous case, when the entire space has the dielectric permittivity equal to that in the wedge $\varepsilon_1=\varepsilon_2\equiv\varepsilon_3$ solving \eqref{eqPhi} leads to
\be
\label{function:angular-divergent}
\Phi_{\nu}^{0}(\theta,\theta')=-{1\over \nu}\left[e^{-\nu|\theta-\theta'|}
+{2\cosh[\nu(\theta-\theta')]\over e^{2\pi\nu}-1}
\right].
\ee
For calculations of the renormalized stress tensor at $\theta=\theta'$ we need to know only the renormalized function
$\Delta \Phi_{\nu}(\theta,\theta')=\Phi_{\nu}(\theta,\theta')-\Phi_{\nu}^{0}(\theta,\theta')$ for $0< \theta,\theta'<\theta_0$:
\be
\label{ftn:angular:renormalized}
\Delta \Phi_{\nu}(\theta,\theta')=-{1\over \nu}\left[\frac{Z}{W} -{2\cosh[\nu(\theta-\theta')]\over e^{2\pi\nu}-1}\right],
\ee
where
\begin{subequations}
\begin{align}
\hspace{-0.3cm}W&= \lambda_{\scriptscriptstyle+++}\cosh{2\pi\nu}
-\lambda_{\scriptscriptstyle+--} \cosh{2\nu(\pi-\theta_0}) \nonumber \\ \label{eqn:W}
&\hspace{2cm}+ \lambda_{\scriptscriptstyle-+-} \cosh{2\nu \theta_{0}} - (\lambda_{\scriptscriptstyle--+}+\lambda_{0}), \\
\hspace{-0.3cm}Z&=\lambda_{\scriptscriptstyle---} \sinh{\nu[\theta+\theta'-2\theta_{0}]} + \lambda_{\scriptscriptstyle+-+} \sinh{\nu[\theta+\theta'-2\pi]} \nonumber \\
&-\lambda_{\scriptscriptstyle-++} \sinh{\nu[\theta+\theta']} -\lambda_{\scriptscriptstyle++-} \sinh{\nu[\theta+\theta'+2(\pi-\theta_{0})]} \nonumber \\
&+\big[\lambda_{\scriptscriptstyle+--} e^{\nu(2\pi-\theta_{0})} - \lambda_{\scriptscriptstyle-+-} e^{-2\nu\theta_{0}} + (\lambda_{\scriptscriptstyle--+}+\lambda_{0}) \nonumber \\
&\hspace{2.55cm}- \lambda_{\scriptscriptstyle+++} e^{-2\pi\nu}\big] \cosh{\nu[\theta-\theta']};
\end{align}
\end{subequations}
above we introduced the notations $\lambda_{0}=8\,\varepsilon_1\varepsilon_2\varepsilon_3$ and $\lambda_{\scriptscriptstyle\pm\pm\pm}=(\varepsilon_1\pm\varepsilon_2)(\varepsilon_1\pm\varepsilon_3)(\varepsilon_2\pm\varepsilon_3)$. Note that the solution for $\theta \in (\theta_{0},\pi)$ can be found from \eqref{ftn:angular:renormalized} by the substitution $\varepsilon_1\to\varepsilon_1$, $\varepsilon_2\to\varepsilon_3$, $\varepsilon_3\to\varepsilon_2$, $\theta_0\to\pi-\theta_0$, $\theta\to\pi-\theta$, and the ensuing replacements in $\lambda$'s.

With the determined Green's function \eqref{transform-inverse:Fourier-Kontorovich-Lebedev}, we are in a position to calculate the stress tensor:
\begin{align}
\label{tensor:energy-stress}
\sigma_{ij}(t_\ins{E},\bs{r},\bs{r}')=-{\varepsilon}\big[ \phi_{;i}\phi_{;j}-{1\over 2}g_{ij}\, \phi_{;k}\phi^{;k}\big];
\end{align}
the metric tensor components in cylindrical coordinates are $g_{00}=-1$, $g_{ij}=g^{ij}=0$ for $i \neq j$, $g^{yy}=1$, $g^{rr}=1$, $g^{\theta\theta}=r^{-2}$. The Fourier transform $\widehat{\sigma}_{ij'}(\zeta_n;\bs{r},\bs{r}')$ of the stress tensor $\sigma_{ij'}(t_\ins{E},\bs{r},\bs{r}')$ is defined similar to \eqref{FT:Green-thermal}. The renormalized Fourier components of the stress tensor $\overline{\sigma}_{ij'}(\zeta_n;\bs{r},\bs{r}')=\widehat{\sigma}_{ij'}(\zeta_n;\bs{r},\bs{r}')-\widehat{\sigma}^{\ins{(div)}}_{ij'}(\zeta_n;\bs{r},\bs{r}')$ can be written in terms of the renormalized Fourier components $\Delta \widehat{G}^{\beta}_{;ij'}= \widehat{G}^{\beta}_{;ij'}-\widehat{G}^{\beta\ins{(div)}}_{;ij'}$ of the thermal Green's function
\begin{subequations}
\begin{align}\label{Dsigmaij}
\overline{\sigma}_{ij'}(\zeta_n;\bs{r},\bs{r}')=&{\varepsilon_3}\big[\Delta \widehat{G}^{\beta}_{;ij'}-{1\over 2}g_{ij'}\,g^{kk'} \, \Delta \widehat{G}^{\beta}_{;kk'}\big] \\
\label{sigma:ES}
+&{1\over 2}g_{ij'}\rho {\partial\varepsilon_{3}\over\partial\rho} \,
g^{kk'}\, \Delta \widehat{G}^{\beta}_{;kk'},
\end{align}
\end{subequations}
so that $\overline{\sigma}_{ij} = g_{j}{}^{j'} \overline{\sigma}_{ij'}$ and then one can take the limit of coincident points  $\overline{\sigma}_{ij}(\zeta_n;\bs{r})=\overline{\sigma}_{ij}(\zeta_n;\bs{r},\bs{r}')|_{\bs{r}=\bs{r}'}$. Here $g_{ij'}(\bs{r},\bs{r}')$ is the operator of parallel transport, which at coincident points reduces to the metric $g_{ij'}(\bs{r},\bs{r}')|_{\bs{r}=\bs{r}'}=g_{ij}(\bs{r})$. Similar to thin films \cite{Dzyaloshinskii:1961} and according to the general theory \cite{Landau:1984}, in a wedge the isotropic elecrostriction part \eqref{sigma:ES} of the stress tensor is absorbed \cite{Note8}\footnotetext[8]{Physically, this absorption follows from the chemical potential, which must be constant for media in equilibrium. One may think of the electrostriction stress as analogous to the gravity in the ocean compensated by the mechanical stresses in water: if it is balanced in the $r$-direction, then due to isotropy it must be balanced in the $\theta$-direction as well. Notably, the $\theta$-dependence of the electrostriction stress leads to non-uniform compression of the matter which is stronger near the interface \cite{Zelnikov:2021}. However, due to isotropy, electrostriction does not contribute to surface tension.}, along with the UV-divergent stress $\sigma^\ins{(div)}_{ij}$ originating from the divergent Green's function \eqref{function:angular-divergent}, by the bare mechanical stress $\sigma^{\ins{(m)}}_{ij}$ to produce the isotropic renormalized mechanical pressure $\sigma^{\ins{(ren)}}_{ij} = - \delta_{ij} p^{\ins{(ren)}}$.

Altogether, the Fourier stress tensor components at $\zeta_n$ are, after raising indices $\overline{\sigma}^{ij} = g^{ik} g^{jl} \overline{\sigma}_{kl}$ to be consistent with the dynamic equations \cite{Note9}\footnotetext[9]{e.g. the Navier-Stokes equations, in which velocity is the contravariant vector; in fact, for direct use in the Navier-Stokes equations the stress tensor components \eqref{eqs:stresses:wedge} need to be transformed to the physical components.},
\begin{subequations}
\label{eqs:stresses:wedge}
\begin{align}
\label{eqs:stress-tt:wedge}
\overline{\sigma}^{\theta\theta} &=\frac{1}{r^5}\!\!\int_0^{\infty}\!\!\d\widetilde{\nu}
\left[2\frac{\partial^2\Delta\Phi_{\nu}}{\partial{\theta}\partial{\theta'}} - (1+2\nu^2)\Delta\Phi_{\nu}\right]_{\theta=\theta'}, \\
\overline{\sigma}^{rr} &=\frac{1}{r^3}\!\!\int_0^{\infty}\!\!\d\widetilde{\nu}
\left[{1\over 2}\Delta\Phi_{\nu}-2\frac{\partial^2\Delta\Phi_{\nu}}{\partial{\theta}\partial{\theta'}}\right]_{\theta=\theta'}, \\
\overline{\sigma}^{yy} &=-\frac{1}{r^3}\!\! \int_0^{\infty}\!\!\d\widetilde{\nu}
\left[{1\over 2}\Delta\Phi_{\nu}+2 \frac{\partial^2\Delta\Phi_{\nu}}{\partial{\theta}\partial{\theta'}} \right]_{\theta=\theta'}, \\
\overline{\sigma}^{r\theta} &=\overline{\sigma}^{\theta r} = -\frac{1}{r^4}\!\! \int_0^{\infty}\!\!\d\widetilde{\nu}
\ 2 \left[\frac{\partial\Delta\Phi_{\nu}}{\partial{\theta}}\right]_{\theta=\theta'},
\end{align}
\end{subequations}
where the measure $\d\widetilde{\nu} = {1 \over  16 \pi}{\d\nu\,\nu\tanh(\pi\nu)}$, and $\Delta\Phi_{\nu}$, $\partial_{\theta}\Delta\Phi_{\nu}$, $\partial_{\theta}\partial_{\theta'} \Delta\Phi_{\nu}$ are computed from \eqref{ftn:angular:renormalized}. Naturally, due to symmetries, $\overline{\sigma}^{y\theta}(\zeta_n;\bs{r})=0$ and $\overline{\sigma}^{yr}(\zeta_n;\bs{r})=0$.

\begin{figure}[t]
\centering \includegraphics[width=80mm]{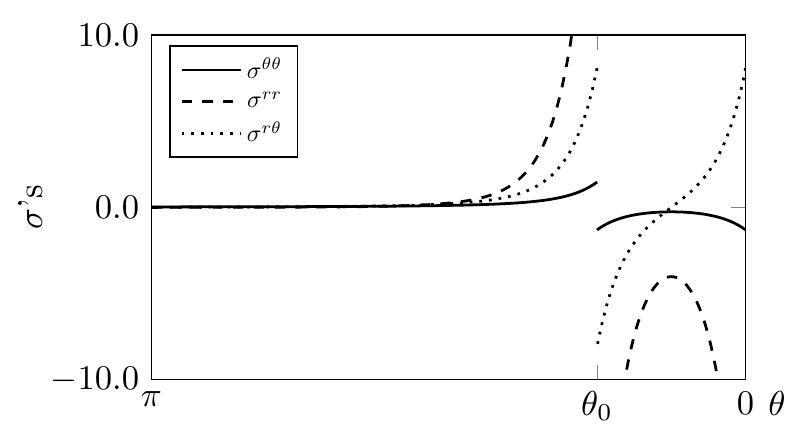}
\caption{\setlength{\rightskip}{0pt} Plots of stress tensor components scaled w.r.t. $16 \pi \beta r^{n}$ with the exponent $n$ corresponding to respective component as per \eqref{eqs:stresses:wedge} at a single frequency and in the case $\theta_{0}=\pi/4$, $\varepsilon_{1}=\varepsilon_{2}=2$, $\varepsilon_{3}=1$, i.e. corresponding to a crack in a dielectric. Integration is performed up to $\nu_{\mathrm{max}}$.} \label{fig:sigmas}\vspace{-0.05cm}
\end{figure}
The classical limit of a slit \cite{Lifshitz:1956}, cf. Fig.~\ref{fig:configurations}a, can be recovered from \eqref{eqs:stresses:wedge} as it corresponds to the region away from the wedge apex $r\to \infty$ and $\theta_0\to 0$, while keeping $\ell=r \theta_0=\const$. The calculations \cite{Brown:1969,Deutsch:1979,Candelas:1982,Bordag:1992,Zelnikov:2021} show that the tangential stress $\sigma_{xx}$ is divergent, cf. Fig.~\ref{fig:configurations}a. For a wedge, we note that divergencies in stresses \eqref{eqs:stresses:wedge} appear already in the normal to the interface stress component $\overline{\sigma}^{\theta\theta}$ if integration is performed w.r.t. $\nu$ to infinity, but in the context of the wedge geometry it is clear that since the interface between media has the thickness $\delta \theta \sim 1/2\nu$ in the angle coordinate and thus the shortest distance to be resolved is intermolecular $a_{0} \sim \delta\theta \, r$, we find $\nu_{\mathrm{max}} \sim r/2 a_{0}$; taking $r \sim 10 \, a_{0}$, we get $\nu_{\mathrm{max}}=5$. Typical stress distributions, clearly demonstrating anisotropy, are shown in Fig.~\ref{fig:sigmas} for the vacuum wedge surrounded by the same dielectric material. The case when the dielectric wedge of the same angle is surrounded by the vacuum is a mirror reflection of Fig. \ref{fig:sigmas} w.r.t. the abscissa.

Next question to consider is on mechanical equilibrium of the wedge. The force density $f^i = \overline{\sigma}^{ik}{}_{;k}$ reads:
\begin{subequations}
\label{eqs:forces}
\begin{align}
\label{eqn:f_theta}
f^{\theta} &= \partial_{\theta}\overline{\sigma}^{\theta\theta}
+\partial_{r}\overline{\sigma}^{\theta r}+{3\over r}\overline{\sigma}^{\theta r}, \\
\label{eqn:f_r}
f^{r} &= \partial_{r}\overline{\sigma}^{rr}+\partial_{\theta}\overline{\sigma}^{\theta r}
+{1\over r}\overline{\sigma}^{rr}-r\overline{\sigma}^{\theta\theta},
\end{align}
\end{subequations}
and, due to symmetry, $f^{y}=0$. In the bulk both $f^{\theta}$ and $f^{r}$ vanish identically. However, $f^{\theta}$, which proves to be independent of $\theta$ outside the interface, if integrated over an elementary volume containing the interface between media $2$ and $3$ yields the force $\d F_\inds{\bot}=r\d F^{\theta}=r^2\d r\d y \int \d\theta f^{\theta}\equiv \d F_\inds{\bot}|_{\theta_0+0}-\d F_\inds{\bot}|_{\theta_0-0}$ acting on every surface element $\d r \d y$ of the interface \cite{Note11}\footnotetext[11]{which is expressed here in an invariant form $\d F_\inds{\bot}=\sqrt{\d F_{\theta}\d F^{\theta}}$, where $\d F_{\theta}=r^2 \d F^{\theta}$}. The corresponding normal pressure $P_\inds{\bot}=\d F_\inds{\bot}/\d r \d y$ tends either to collapse or unfold the wedge, cf. Fig.~\ref{fig:configurations}b:
\begin{align}
P_\inds{\bot}=\frac{8}{r^3}\!\int_0^{\infty}{\d\widetilde{\nu}\nu  \frac{\lambda_{\scriptscriptstyle+--} \sinh{2\nu(\theta_{0}-\pi)} - \lambda_{\scriptscriptstyle-+-} \sinh{2\nu\theta_{0}}}{W}}. \nonumber
\end{align}
This pressure is finite because the divergencies on either side of the interface have opposite signs; hence, $P_\inds{\bot}$ is independent of the cut-off $\nu_{\mathrm{max}}$! First, a few clarifications about the sign of $P_\inds{\bot}$, which can be illustrated using the asymptotics of the integrand $\mathcal{I}$ in the expression for $P_\inds{\bot}$:
\begin{subequations}
\label{integrand}
\begin{align}
\label{integrand:0}
\theta_{0} \rightarrow 0: \ \mathcal{I} \sim \frac{(\varepsilon_{1}-\varepsilon_{3})(\varepsilon_{3}-\varepsilon_{2})}{2(\varepsilon_{1}+\varepsilon_{2}) \varepsilon_{3}}; \\
\label{integrand:pi}
\theta_{0} \rightarrow \pi: \ \mathcal{I} \sim \frac{(\varepsilon_{1}-\varepsilon_{2})(\varepsilon_{3}-\varepsilon_{2})}{2(\varepsilon_{1}+\varepsilon_{3}) \varepsilon_{2}}.
\end{align}
\end{subequations}
As we know from the planar geometry case \cite{Lifshitz:1956,Dzyaloshinskii:1961,Lifshitz:1980}, the force between two dielectric media $\varepsilon_{1}$ and $\varepsilon_{2}$ separated by the vacuum $\varepsilon_{3}=1$ corresponding to the limit \eqref{integrand:0} is $P_\inds{\bot}<0$ implying attraction between the two dielectrics. The other limit \eqref{integrand:pi} can be verified with a liquid helium film on glass, $\varepsilon_{1} > \varepsilon_{2}$, which leads to $P_\inds{\bot}<0$ corresponding to repulsion in this case, consistent with the tendency of the liquid helium film to thicken \cite{Dzyaloshinskii:1961}.

\begin{figure}[t]
\centering \includegraphics[width=80mm]{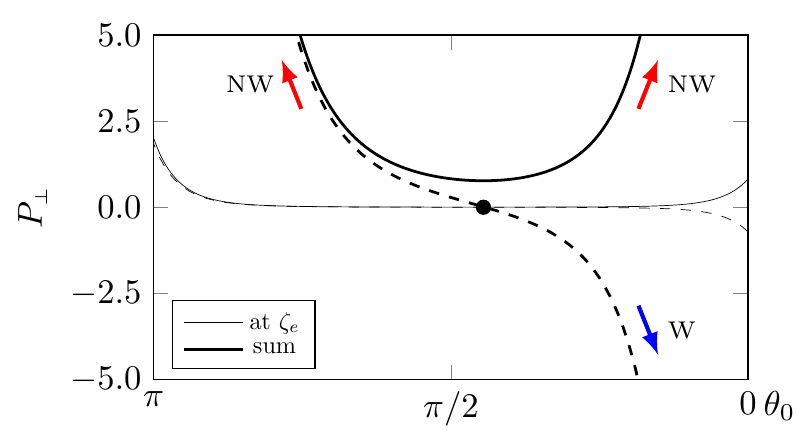}
\caption{\setlength{\rightskip}{0pt} Plot of $P_\inds{\bot}$ for water shown at a single absorption frequency $\zeta_{e}$ and summed up over all Matsubara frequencies $\zeta_{n}$ using the corresponding $\varepsilon(\i \zeta_{n})$ dependence \cite{Israelachvili:2011}: solid curves corresponds to water on mica and dashed ones to carbon disulfide on teflon. Red arrows -- non-wetting (NW) tendency to turn away from $\theta_{0}=0$ towards $\pi$. Blue arrow -- wetting (W) tendency $\theta_{0} \rightarrow 0$.} \label{fig:deltaH}
\end{figure}

An example computation of $P_\inds{\bot}$ is demonstrated in Fig. \ref{fig:deltaH} for (a) water on mica, which is known to wet perfectly, $\theta_{0}=0-5^{\circ}$, and (b) carbon disulfide on teflon, which on macroscopic scale shows contact angle of $80^{\circ}$. Let us first consider implications of the computed $P_\inds{\bot}$ in isolation from the surface tension effects, which is possible since $P_\inds{\bot}$ does not account for any contributions to interfacial tensions as in the sharp interface formulation considered here, due to antisymmetry of the stresses, the latter give zero contribution to these tensions \cite{Candelas:1982,Zelnikov:2021}. In the case (a) the fact that $P_\inds{\bot}$ is positive and non-zero implies that there is no mechanical equilibrium (angle $\theta_{0}$) -- the wedge interface tends to turn to $\theta_{0}=\pi$. In the case (b) there is an equilibrium angle, but it is obviously unstable, so the contact angle $\theta_{0}$ may collapse either to $0$ or $\pi$ as dictated by the minimum of potential energy including not only the surface tension energy, but also the energy of van der Waals stress field. The same behavior as that for water on mica is exhibited for water on PVC, which is known to be non-wetting, and for benzene on fused quartz, which exhibits the contact angle $\theta_{0}=11^{\circ}$. In the latter case the liquid is non-polar and hence the Lifshitz theory accounting for London forces only should be more accurate, though it is still widely applied even to polar liquids such as water \cite{Israelachvili:2011}; however, one can anticipate that due to polarity of water molecules, which leads to strong hydrogen bonds, and the Keesom effect dominating that of London, the deviations from the Lifshitz theory should be significant \cite{Zelnikov:2021}. Both generic -- parabola up and cotangent-like curves shown in Fig.~\ref{fig:deltaH} -- demonstrate either perfect wetting or non-wetting: even if equilibria exist, they prove to be unstable. The only stable situation would be possible if the cotangent-like curve in Fig.~\ref{fig:deltaH} is mirror reflected to become tangent-like. However, as follows from the asymptotics \eqref{integrand}, for a typical wetting situation when $\varepsilon_{2} \approx 1$ this would require $\varepsilon_{1}>\varepsilon_{3}$ thus violating the Lifshitz limit at $\theta_{0}\rightarrow\pi$. Hence, for all liquid-on-solid wetting situations with air being phase 2, there is no stable contact angle $\theta_{0}$ other than $0$ or $\pi$, should one focus on the force $P_\inds{\bot}$ alone.

However, if one considers a liquid wedge on a solid substrate, then it is known from classical macroscopic considerations that minimization of the sum of energies of all interfaces leads to the Young equation \cite{Young:1805}, $\gamma_{21}-\gamma_{31}=\gamma_{23} \cos{\theta_{0}}$, that can be viewed as the projection of surface tension forces (Young force diagram) on the substrate plane. This equation, though, does not account for the bulk van der Waals stresses, which makes Young's equation inapplicable near the wedge corner in the same way as one would not apply it to solids, where internal stresses play the dominant role. Therefore, as opposed to the commonly used Frumkin-Derjaguin approach \cite{Frumkin:1938,Derjaguin:1940}, in which Young's equation stays unmodified in the presence of disjoining pressure, the correct force balance should add the projection of the resultant van der Waals force acting on the $2-3$ interface $-\langle P_\inds{\bot}\rangle \cos{\theta_{0}}$ to Young's equation. Since surface tension forces applied to flat interfaces in the Young force diagram are independent of the distance $r$ to the wedge tip, at sufficiently short distances they are dominated by $\d F_\inds{\bot} \sim r^{-3}$, which tend to turn the wedge interface either toward $\theta_{0}=0$ or $\pi$. When the non-zero width of interface $2-3$ is taken into account, this happens at the distances close to the interface thickness, i.e. on the order of a few molecular distances $a_{0}$, where van der Waals stresses and their part contributing to surface tension become inseparable \cite{Zelnikov:2021}. Therefore, the present study establishes that the contact angle $\theta_{0}$ reported in literature from macroscopic observations is different from the actual one, at which the interface meets the substrate, and instead is set asymptotically at the distances $r$ where surface tension effects become dominant thus leading to the classical Young force diagram. Therefore, the wedge interface 2-3 must necessarily be curved, which also follows from the nonuniformity of pressure $P_\inds{\bot}$ along the interface.

The presented theory is also applicable in the case of a wedge dynamically moving with velocity $U$ along the substrate, i.e. the moving contact line problem. Clearly, the viscous stresses $\sim \mu U/r$ are on the order of the computed here Derjaguin's stresses at $r^{*} \sim \sqrt{A_{\mathrm{H}} / 6 \pi \mu U} \sim 100 \, \mathrm{nm}$, where $A_{\mathrm{H}}$ is taken for water on mica. Below this scale, the Derjaguin stresses dominate due to $r^{-3}$ divergence. Effectively, this means that for $r<r^{*}$ the liquid cannot be considered as Newtonian with the same bulk viscosity $\mu$ as that for $r>r^{*}$. The increased, but non-divergent, stresses for $r<r^{*}$ enable ripping of liquid from the substrate in the case of a receding contact line.

\end{document}